\documentclass[10pt,letterpaper]{article}
\usepackage{fullpage}

\usepackage[utf8]{inputenc}
\usepackage[english]{babel}

\usepackage{graphicx}
\usepackage{gensymb}

\usepackage[labelformat=simple,font=small]{subcaption}
\usepackage[font=small]{caption}
\addto\captionsenglish{}

\usepackage[intlimits]{amsmath}

\usepackage{cite}

\newcommand{\Var}{\mathrm{Var}}

\let\oldhat\hat
\renewcommand{\hat}[1]{\oldhat{\mathbf{#1}}}

\title{Measurement of particle motion in optical tweezers embedded in a Sagnac interferometer}
\date{}

\usepackage{authblk}

\author[1,2]{Ivan Galinskiy \footnote{Corresponding author. Email address: \texttt{iv.galinskiy@ciencias.unam.mx}}}
\author[1,3]{Oscar Isaksson}
\author[1,4]{Israel Rebolledo Salgado}
\author[2]{Mathieu Hautefeuille}
\author[1]{Bernhard Mehlig}
\author[1]{Dag Hanstorp}

\affil[1]{Department of Physics, University of Gothenburg, SE-412 96 Gothenburg, Sweden}
\affil[2]{Facultad de Ciencias, Departamento de Fisica, Universidad Nacional Autonoma de Mexico, Ciudad Universitaria, DF 04510, Mexico}
\affil[3]{Department of Applied physics, Chalmers University of Technology, SE-412 96 Gothenburg, Sweden}
\affil[4]{Instituto de Ciencias Físicas, Universidad Nacional Autonoma de Mexico, 62210, Cuernavaca, Mexico}

\begin{document}
\maketitle

\begin{abstract}
 We have constructed a counterpropagating optical tweezers setup embedded in a Sagnac interferometer in order to increase the sensitivity of position tracking for particles in the geometrical optics regime. Enhanced position determination using a Sagnac interferometer has previously been described theoretically by Taylor \textit{et al.} [Journal of Optics 13, 044014 (2011)] for Rayleigh-regime particles trapped in an antinode of a standing wave. We have extended their theory to a case of arbitrarily-sized particles trapped with orthogonally-polarized counterpropagating beams. The working distance of the setup was sufficiently long to optically induce particle oscillations orthogonally to the axis of the tweezers with an auxiliary laser beam. Using these oscillations as a reference, we have experimentally shown that {Sagnac-enhanced} back focal plane interferometry is capable of providing an improvement of more than 5 times in the signal-to-background ratio, corresponding to a more than 30-fold improvement of the signal-to-noise ratio. The experimental results obtained are consistent with our theoretical predictions. In the experimental setup, we used a method of optical levitator-assisted liquid droplet delivery in air based on commercial inkjet technology, with a novel method to precisely control the size of droplets. 
\end{abstract}

\section{Introduction}
The techniques of optical levitation and optical tweezers were introduced by Ashkin \textit{et al.} in 1971 and 1986, respectively \cite{ashkin-levitation,ashkin-tweezers}. In optical levitation, a weakly focused, vertically aligned laser beam creates an optical pressure force on a dielectric particle which can become sufficiently strong to compensate gravity. A weak gradient force provides stability in the horizontal plane, creating a stable trap \cite{ashkin-levitation-stability}. Optical levitation does not require high numerical apertures (NA) for trapping. Its main advantage, thus, is the large working distance of the focusing optics. This property has been used, for example, to determine the electric charge of levitated oil droplets \cite{oscar}.

In single-beam optical tweezers, on the other hand, a tightly focused beam is used to create a purely optical three-dimensional trap that, unlike the case of optical levitation, does not require any external forces for stability \cite{ashkin-tweezers}. This has been of great importance to the biomedical sciences, where non-contact individual-cell manipulation experiments were made possible, such as adaptive cell sorting \cite{sorting}. The great advantage with the optical tweezers is that the same objective can be used for both trapping and imaging of the investigated object. Hence, the optical tweezers can be incorporated in any type of advanced microscope \cite{xi,multiphoton}. The optical tweezers has then evolved into a standard instrument in microbiology \cite{svoboda}. The main limitation of single-beam optical tweezers is the high numerical aperture required for trapping, which severely reduces the working distance of the focusing microscope objective or lens \cite{air-trapping}.

An optical tweezers setup can also be \emph{counterpropagating}, i.e. composed of two beams that propagate in opposite directions. Several implementations of such setups exist. The most commonly used implementation of this method consists of two opposite-travelling trapping beams focused into the same focal point by a pair of lenses \cite{counter-propagating}. This arrangement permits to reduce the required NA for successful trapping and thus, increase the working distance of the focusing optics \cite{3d-manipulation}. This is the implementation that will be considered in the rest of this work. Alternatively, fiber-based counterpropagating tweezers have been realized, where two optical fibers were facing each other at a distance of a few hundred micrometers \cite{fiber-tweezers}. This arrangement has been used, for example, to non-destructively stretch individual living cells \cite{stretching}.

One of the most important properties of optical tweezers is that they offer the possibility of precisely measuring the position of a trapped particle. For example, the stepping motion of a single molecule could be resolved into individual steps by integrating a Wollaston prism-based interferometer into an optical tweezers \cite{kinesin}. The method that is most commonly used for tracking the position of a particle in optical tweezers is the back focal plane (BFP) interferometry in which the back focal plane of the condenser lens is imaged onto a quadrant photodiode detector (QPD) \cite{qpd}. While the back focal plane interferometry allows for subnanometer resolution in the tracking of micron-scale particles \cite{resolution}, it is subject to shot noise as any other optical method of measurement. This ultimately limits the resolution of position measurements for particles of any size. The limiting effects of shot noise on position measurement of Rayleigh-regime particles, i.e. particles much smaller than the wavelength of the trapping light, were shown theoretically by Taylor \textit{et al.} \cite{sagnac}. There, the particle was assumed to be trapped in an antinode of the standing wave produced by same-polarization counterpropagating beams.

By embedding a counterpropagating setup into a Sagnac interferometer, it is possible to selectively attenuate the trapping field, while preserving the asymmetrical field components that carry information about the position of the particle. This increases the signal-to-background ratio and consequently, the signal-to-noise ratio, as shown theoretically in \cite{sagnac} by Taylor \textit{et al.} for the case of Rayleigh scatterers.

There are many applications of counterpropagating optical tweezers that use particles with sizes comparable to or larger than the wavelength of trapping light, being either in the Mie or the geometrical optics regimes \cite{kinesin,cooling}. There, the trapping is normally achieved with orthogonally polarized counterpropagating beams to avoid the formation of a standing wave that would lead to a periodically varying trapping potential. With minor modifications, this class of applications could benefit from the increase of position measuring sensitivity offered by embedding them into a Sagnac interferometer. To address this possibility, we generalized the theory by Taylor \textit{et al.} \cite{sagnac} to arbitrary-size spherical particles in a setup where the counterpropagating beams are \emph{orthogonally} polarized. After that, we confirmed the enhanced signal-to-noise ratio experimentally, continuing our previous work \cite{laop}. The reflections from the particle itself were shown to be detrimental to the signal-to-noise ratio by Taylor \textit{et al.} \cite{sagnac}. In our setup they are successfully eliminated by polarization filtering in the interferometer.

As a second aspect, we implemented a technique for high-efficiency, single-droplet generation in air based on a commercial inkjet cartridge \cite{inkjet} with a novel evaporation-based diameter adjustment method that allowed us to create droplets with precise control of their size. The trapping was greatly simplified by using levitator-assisted particle delivery into the optical tweezers \cite{inkjet-levitation,levitator-feeding}. This allowed us to create and trap droplets one-by-one, in contrast to nebulizer-based methods \cite{solid-air-trapping}. The long working distance (WD) of the aspherical lenses used for the optical tweezers simplified optical access to the trapped particle from a direction perpendicular to the optical axis of the tweezers. This allowed us to optically induce transverse oscillations of the trapped particle by applying a modulated ``side-pressure'' laser beam aligned orthogonally to the trapping beams. These particle oscillations served as a reference for signal-to-background measurements. The obtained experimental results closely matched our theoretical predictions.

\section{Experimental setup}

\subsection{Optical levitator}
\begin{figure}[htb]
  \centering  
  \includegraphics[width=0.85\textwidth]{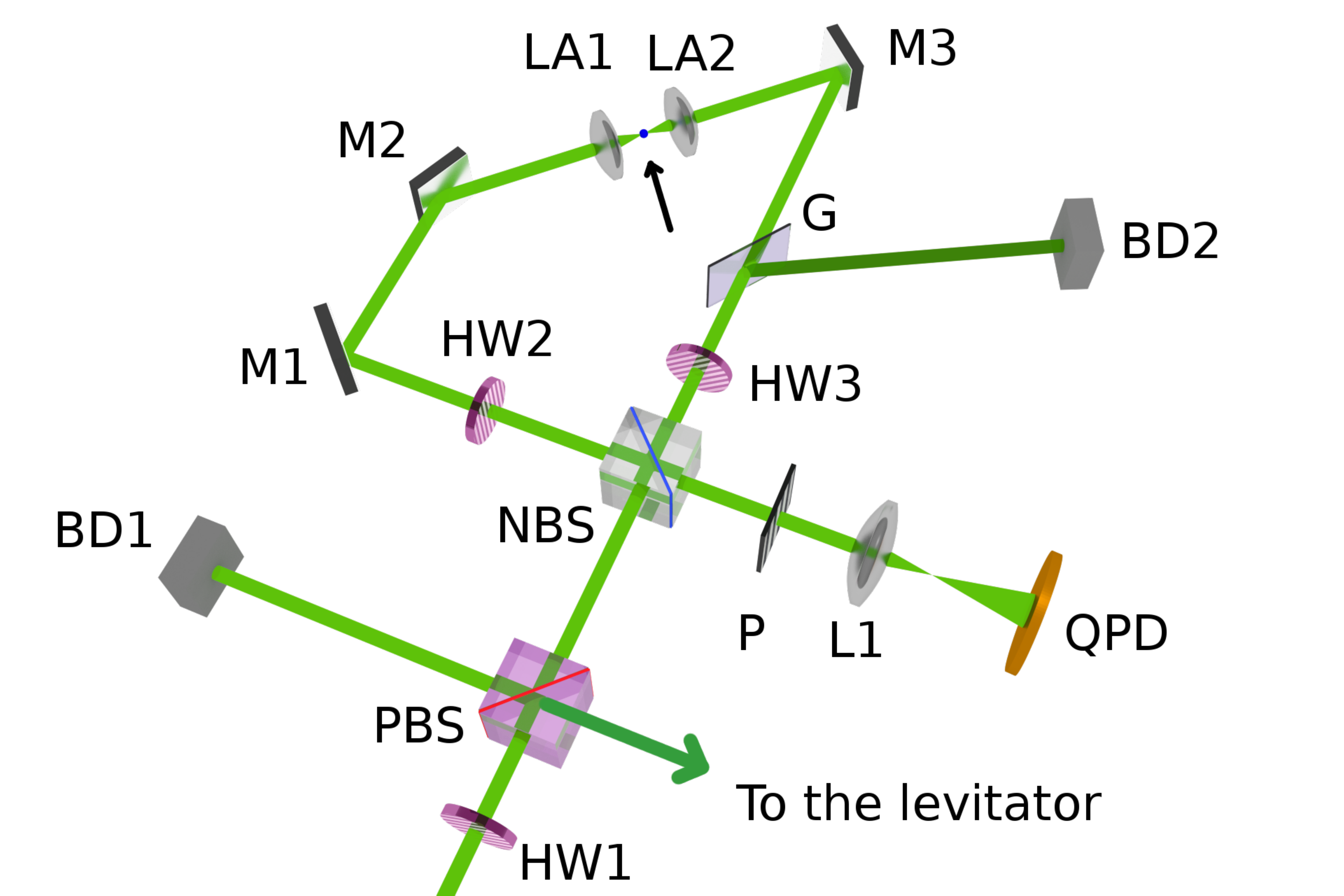} 
  \caption{The optical setup of the optical tweezers subsystem. After passing through the HW1 half-wave plate and the polarizing beamsplitter PBS, the incoming light beam is split, so that chosen parts of it go into the optical levitator subsystem (not shown) and into the optical tweezers subsystem. The latter beam is then split by the non-polarizing beamsplitter NBS into two counterpropagating beams. The HW3 half-wave plate rotates the polarization of the counterclockwise beam (CCW) to guarantee orthogonal polarizations between the CW and the CCW beams in the trapping point (marked by an arrow) between the aspherical lenses LA1 and LA2. The glass plate G serves to selectively attenuate the CCW beam, redirecting a part of it onto the beam dump BD2. The CW and CCW beams recombine at the NBS beamsplitter and the interference pattern is filtered by the polarizer P to remove any reflections from the optical system. The lens L1 focuses the back focal plane of LA1 and LA2 onto the quadrant photodiode detector QPD. At the same time, the light returning back to the laser is reflected by PBS onto BD1. Additionally, HW2 compensates for the phase shift between the orthogonal polarizations, caused by reflections from the mirrors M1, M2 and M3.}
  \label{fig:setup}
\end{figure}

The light in the optical setup is generated by a DPSS green laser module (Laser Quantum, 532 nm, 2 W maximum power). The output beam is split by a polarizing beamsplitter into two beams (PBS in Fig. \ref{fig:setup}), where the split ratio is controlled by rotating the half-wave plate HW1. The transmitted part of the beam is guided into the optical tweezers system shown in Fig. \ref{fig:setup}, while the reflected part is guided into the optical levitation system, entering from below in Fig. \ref{fig:3dcell}.

\begin{figure}[htb]
\centering
  \begin{subfigure}[t]{0.3\textwidth}
    \includegraphics[width=\textwidth]{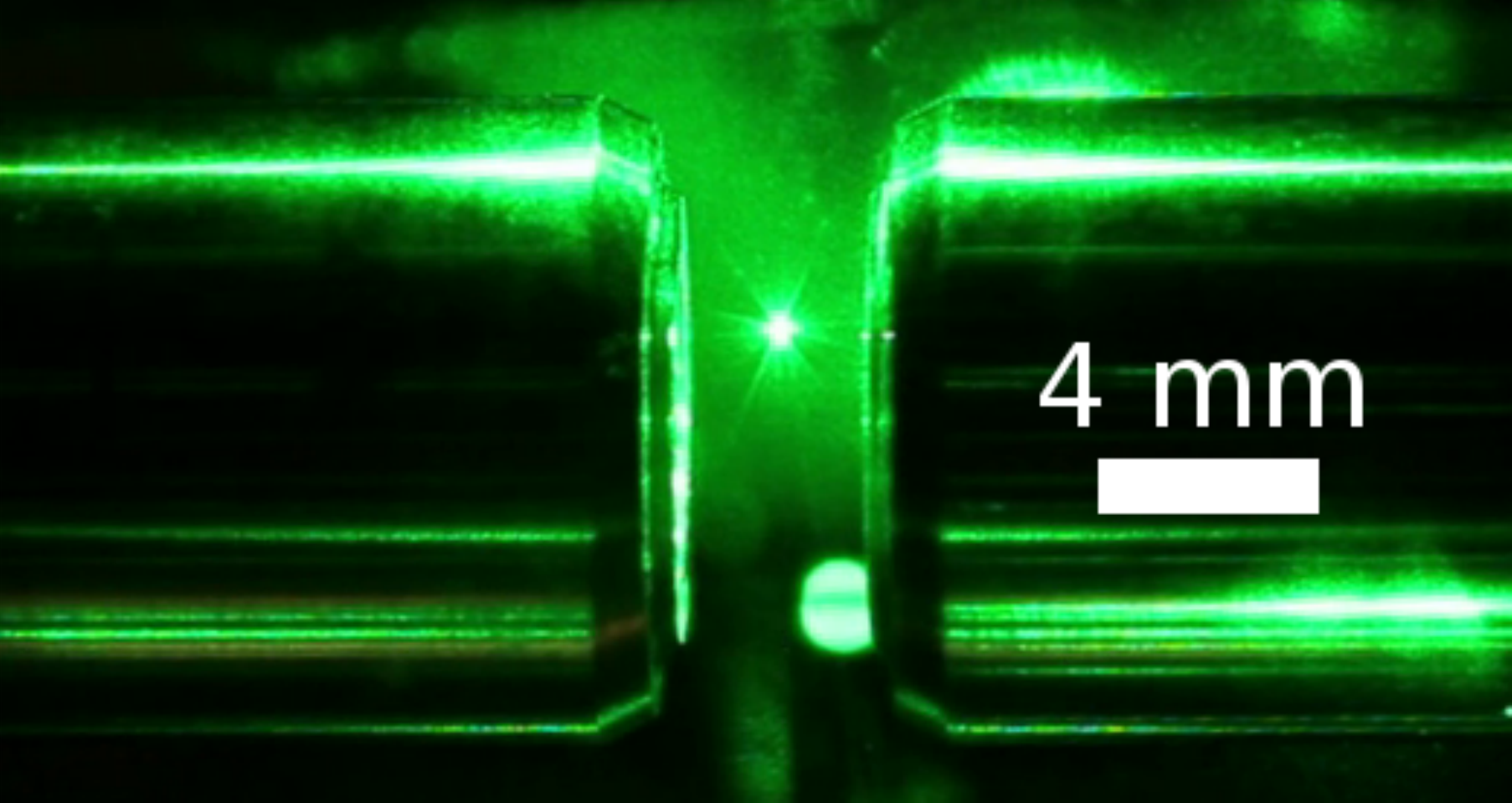}
    \caption{Particle just trapped in the levitator.}
    \label{fig:tw_seq1}
  \end{subfigure}
  \quad
  \begin{subfigure}[t]{0.3\textwidth}
    \includegraphics[width=\textwidth]{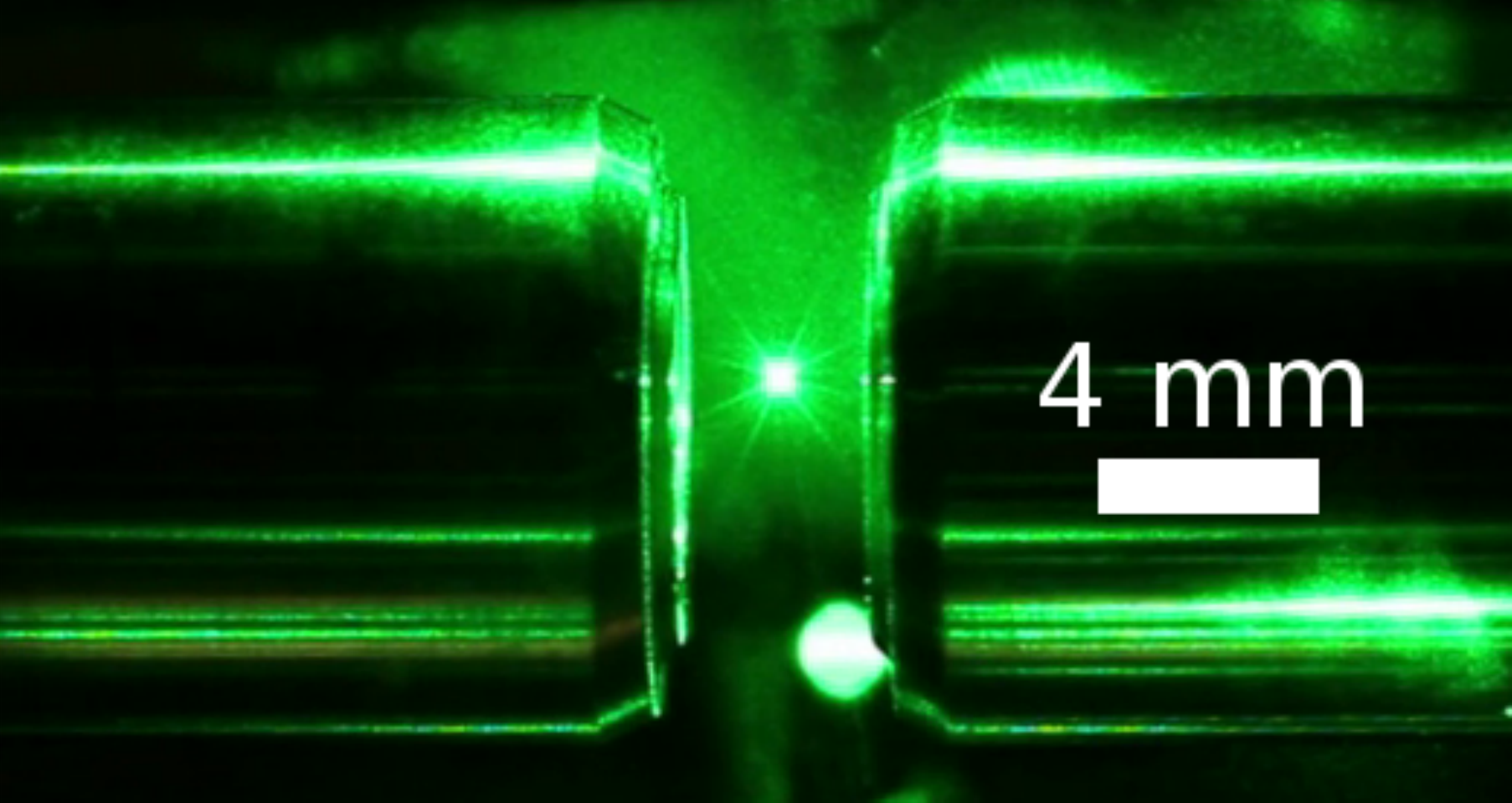}
    \caption{Particle being lowered by decreasing the levitator power.}
    \label{fig:tw_seq2}
  \end{subfigure}
  \quad
  \begin{subfigure}[t]{0.3\textwidth}
    \includegraphics[width=\textwidth]{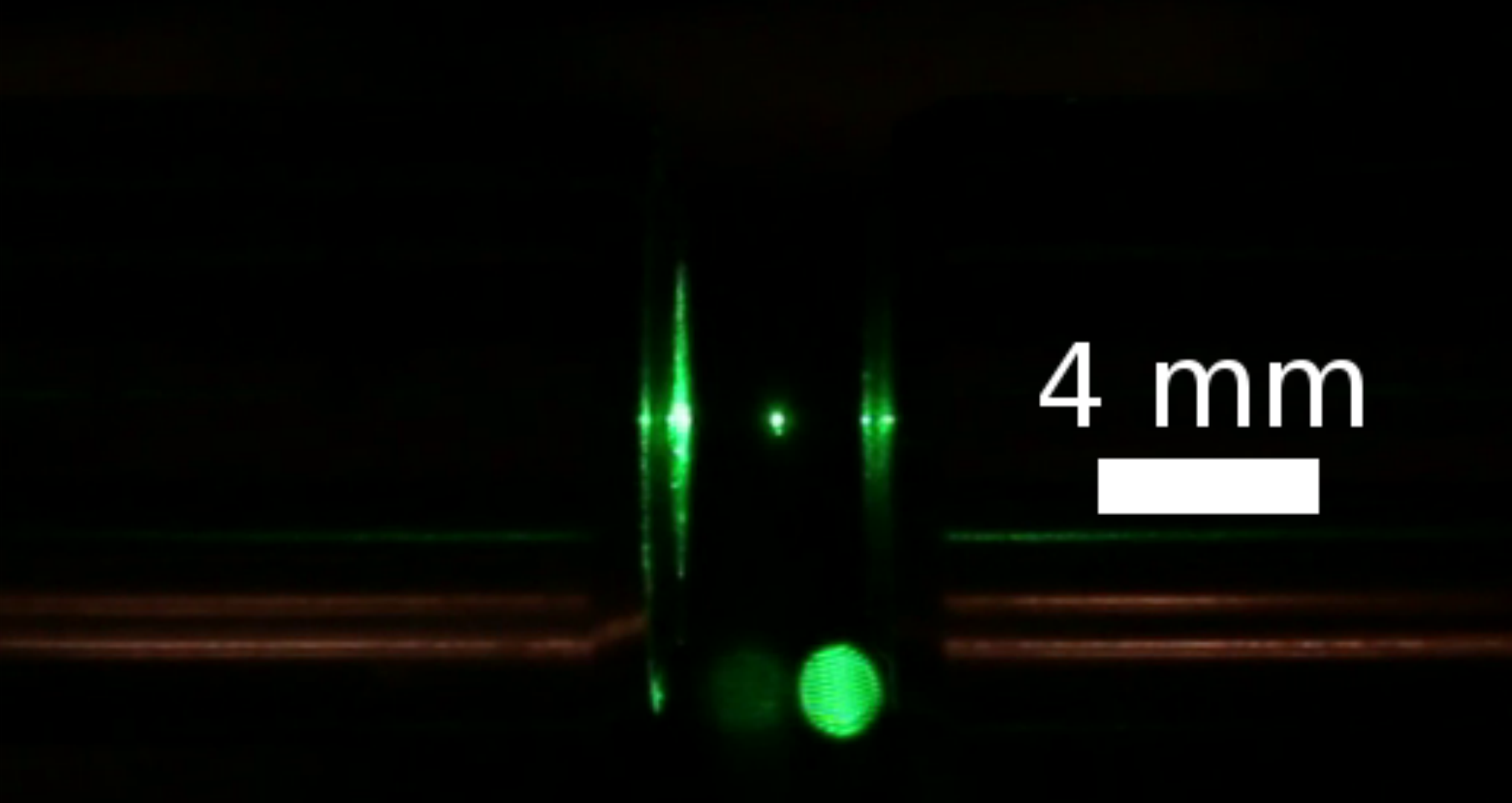}
    \caption{Particle trapped in the counterpropagating optical tweezers with the levitator fully attenuated.}
    \label{fig:tw_seq3}
  \end{subfigure}
  \caption{Photographs of successive steps of trapping a single levitated glycerol particle. Aspherical lenses (not visible) are mounted at the inner ends of the left and the right metallic cylinders.}
  \label{fig:tw_seq}
\end{figure}

The vertically directed optical levitator beam (marked as ``LEV'' on Fig. \ref{fig:3dcell}), is focused by a lens with a focal length of 100 mm. The optical setup is configured to place the focal point of the levitator beam about 2 mm below the optical tweezers' trapping point (Fig. \ref{fig:3dcell}). With this arrangement, the stable trapping point of the levitator (using 1000 mW of laser power) is localized approximately 2 mm above the focal point of the optical tweezers (Fig. \ref{fig:tw_seq1}). The waist diameter of the Gaussian levitation beam was calculated to be 17 $\mu$m.

\begin{figure}[htb]
  \centering  
  \includegraphics[width=0.4\textwidth]{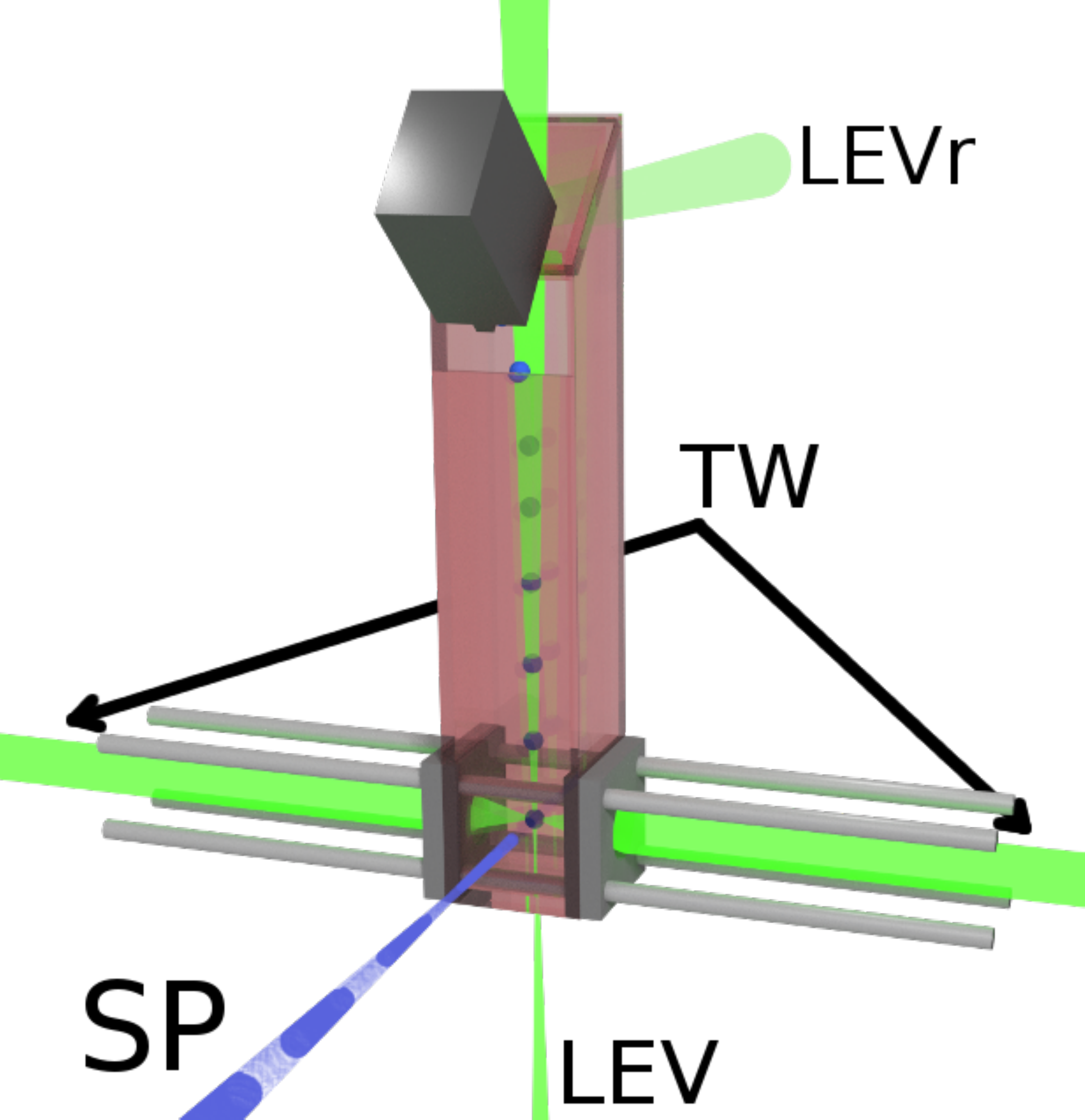}
  \caption{A schematic close-up of the trapping point together with the glass cell. The cartridge (gray box) generates droplets that descend vertically through the levitation beam (LEV) until they are trapped in the counterpropagating optical tweezers beams (TW), where they are perturbed horizontally by a TTL-modulated side-pressure laser (SP). The top glass plate is tilted to direct the Fresnel reflections (LEVr) out of the cell.}
  \label{fig:3dcell}
\end{figure}

That way, it is possible to modify the vertical position of the particle by changing the light power coupled into the levitator, i.e. by changing the split ratio for the PBS beamsplitter. The power in the levitator is then reduced, which leads to an increase of the power in the optical tweezers. This effectively lowers the droplet until it descends into the trapping point of the counterpropagating optical tweezers (Fig. \ref{fig:tw_seq2}), where it is stably trapped. After that, all the light is coupled into the optical tweezers (Fig. \ref{fig:tw_seq3}).

\subsection{Optical tweezers}
The p-polarized beam from the laser enters the non-polarizing beamsplitter NBS where it is split into one clockwise (CW) and one counterclockwise (CCW) beam. The CCW beam becomes s-polarized after passing through the half-wave plate HW3, whose axis is rotated at 45$\degree$ to the horizontal plane. After that, the glass plate G, tilted at approximately Brewster's angle, reflects a part of the CCW beam, which is dumped in the beam dump BD2. This is used to compensate for any power imbalance between the CW and CCW beams caused by the NBS beamsplitter (Thorlabs BSW26). Then, after reflecting on the mirror M3, the beam passes through the aspherical lens LA2 (Thorlabs/Geltech C330TMD-A), where it is focused onto the particle. The lenses LA1 and LA2 are aligned to make their focal points coincide in a single spot, which will be the trapping position for the particle. After passing through the particle, the beam is collimated again by the lens LA1 (Thorlabs/Geltech C330TMD-A). The focused cone of light has a numerical aperture of 0.30.

The CW beam, on the other hand, passes through the half-wave plate HW2 (that does not alter the beam's polarization), the aspherical lenses and then through the glass plate G. Since this beam is p-polarized, and the angle of G is close to Brewster's angle, only a slight fraction of this beam is reflected. In this way both beams have the same power not only at the trapping point, but also when "meeting" at NBS. Finally, the CW field passes through HW3 which converts it from p-polarized to s-polarized. 

Therefore, both the CW and CCW beams are s-polarized when combined again at NBS. Note that the fast axis of the half-wave plate HW2 is horizontal and thus, the waveplate simply compensates for the additional $\pi$ phase shift between the different polarizations caused by the mirrors and the HW2 waveplate. Otherwise, the additional $\pi$ phase shift between the beams would lead to an interchange between the location of the light port with the dark port of the interferometer.

After recombining at NBS, the beams interfere constructively when travelling back towards the laser and interfere destructively when travelling towards the quadrant photodiode detector (QPD), thus creating the so-called ``dark port'' of the interferometer. There, since all the reflected light returning to the NBS beamsplitter is p-polarized, a polarizer (P) eliminates nearly all the light that was reflected, either from the surface of the particle or from the optical components of the setup. Finally, a lens (L1) images the back focal planes of both lenses onto the QPD.

A Sagnac interferometer usually creates a severe backreflection problem because of constructive interference in the beam returning to the laser. However, in this setup the backreflected beam is polarized orthogonally to the input beam, and it is thus possible to filter all the backreflected light in the PBS polarizing beamsplitter, directing it to the beam dump BD1 instead.

In order to compare the performance of the Sagnac-enhanced position sensing to the usual back focal plane interferometry method, we take advantage of a ghost reflection from the NBS beamsplitter. This ghost reflection is a highly attenuated sample of the \emph{CW-propagating beam only}, which corresponds to traditional BFP measurements. Additionally, this reflection propagates at a slight angle ($\approx 0.5 \degree$) to the main beam (``dark port beam''), which makes it possible to separate it. Therefore, by blocking the dark port beam and opening the ghost reflection beam (or vice versa), it is possible to conveniently ``switch`` between the two modes of operation (traditional BFP and Sagnac-enhanced BFP), while using the same imaging lens (L1).

\subsection{Droplet generation using an inkjet cartridge}

\begin{figure}[htb]
\centering
\includegraphics[width=0.6\textwidth]{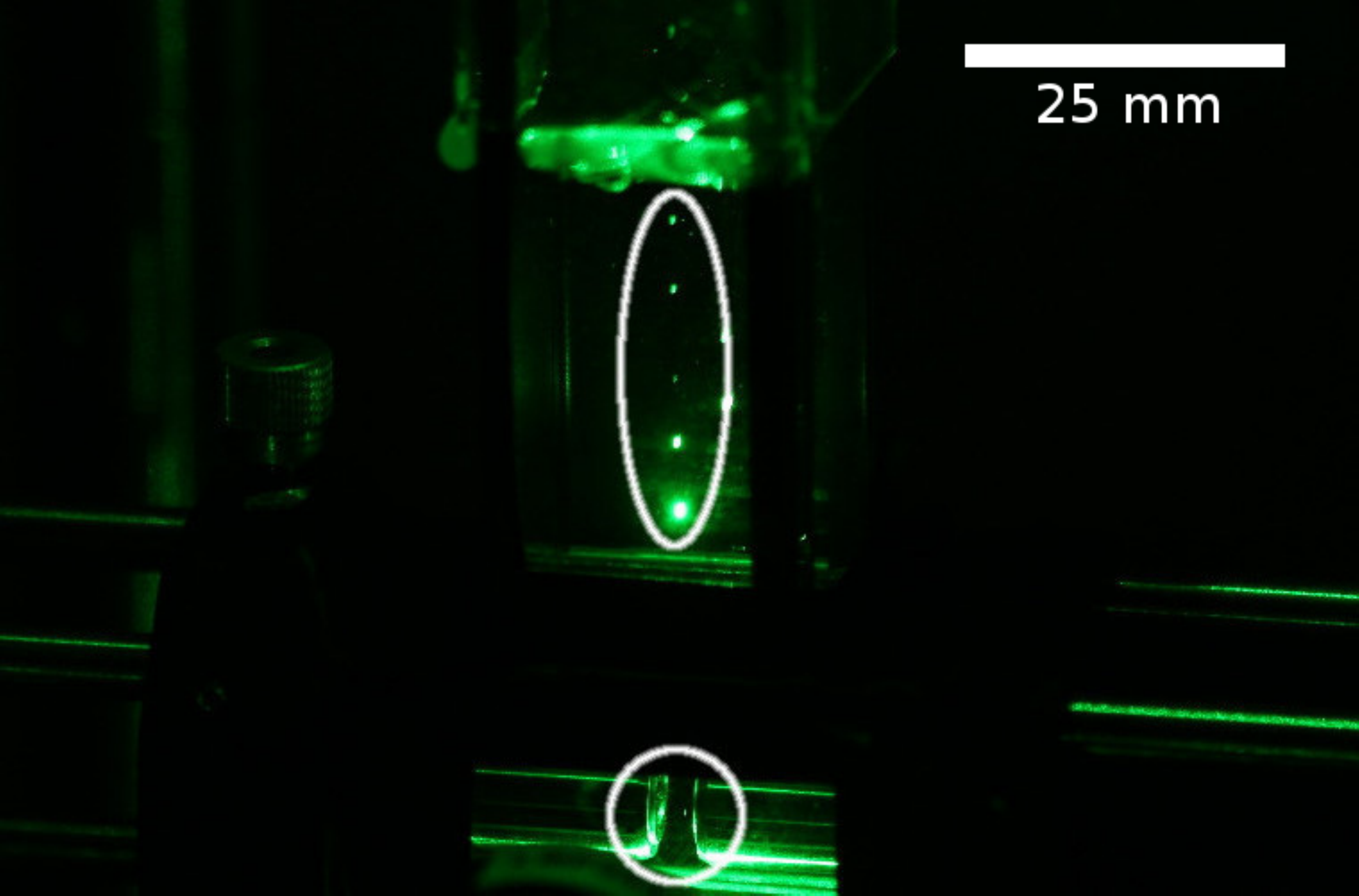}
\caption{A photograph of the trapping cell during operation. Droplets are continuously generated by the inkjet cartridge (not shown). Several droplets (bright dots in the upper ellipse) can be seen descending through the levitator beam towards the trapping point of the optical tweezers (inside the lower circle).}
\label{fig:cell_real}
\end{figure}

We have constructed a system for generating droplets of controllable size using a commercial inkjet cartridge (Hewlett-Packard C6614). The system is based on some of the methods demonstrated in \cite{inkjet,inkjet-levitation}. Our technique of droplet size control consists of filling the cartridge with a 5\% solution of glycerol in water. While the diameter of the particles emitted by the cartridge was measured to be a nearly constant 34 $\mu$m \cite{inkjet}, the quick evaporation of water from the falling droplets reduces the diameter to $11 \pm 1$ $\mu$m in average (the measurement procedure is discussed in section \ref{sec:diameter}). This size is more suitable for trapping due to the lower mass, while still being much larger than the wavelength of trapping light.

The cartridge is positioned out of line of the vertical levitation laser beam in order to avoid the high-intensity light from reaching the cartridge surface, which would create serious convection flows in the cell and damage in the cartridge itself. In order for the droplets to reach the levitator beam, the cartridge is tilted at an angle of approximately 45$\degree$ with respect to the horizontal plane (Fig. \ref{fig:3dcell}). In this geometry, the ejected droplets fly 3-5 cm at an angle of 45$\degree$ before reaching their terminal velocity, which allows them to reach the levitation beam and descend vertically afterwards (Fig. \ref{fig:cell_real}). With proper protection against air currents, the droplets fall predictably into the focal region of the levitation beam where the optical pressure is sufficient for stable levitation.

\subsection{Trapping cell}
The main purpose of the trapping cell is to shield the droplets from external air currents. Additionally, it allows the levitation light beam to pass through the setup without reflecting it back into the cell, which would create convection flows. This is achieved by tilting the upper glass plate 45$\degree$ (Fig. \ref{fig:3dcell}) to allow the Fresnel reflections (marked as ''LEVr`` in Fig. \ref{fig:3dcell}) to be deflected, and placing a beam dump outside the cell. The cell is sufficiently tall to guarantee the full evaporation of the solvent (water) in the particles, allowing them to achieve the desired diameter prior to trapping. The trapping cell is constructed with microscope slides supported by the optical cage system that is also holding the lens assembly, as shown in Figs. \ref{fig:3dcell} and \ref{fig:cell_real}.

\subsection{Horizontal pressure laser}

The system to induce horizontal oscillations of the trapped particle is conceptually similar to the optical levitator described above. A separate DPSS laser (532 nm, 200 mW) is controlled by a Pulse Width Modulated (PWM) signal with a given frequency (0-10 kHz) and a controllable duty cycle. Its output is steered and focused horizontally onto the trapped particle (marked as ''SP`` in Fig. \ref{fig:3dcell}). The beam waist is deliberately located slightly behind the particle in order to reduce the total optical force produced by the side beam. The Airy-like diffraction pattern produced after passing through the trapping point with a trapped particle (Fig. \ref{fig:airy}) indicates that the beam is incident on the particle, thus confirming the correct alignment.

The oscillations induced by this system are monitored using a quadrant photodiode detector (QPD in Fig. \ref{fig:setup}).

\subsection{Particle size measurement}\label{sec:diameter}

\begin{figure}[htb]
  \centering  
  \includegraphics[width=0.4\textwidth]{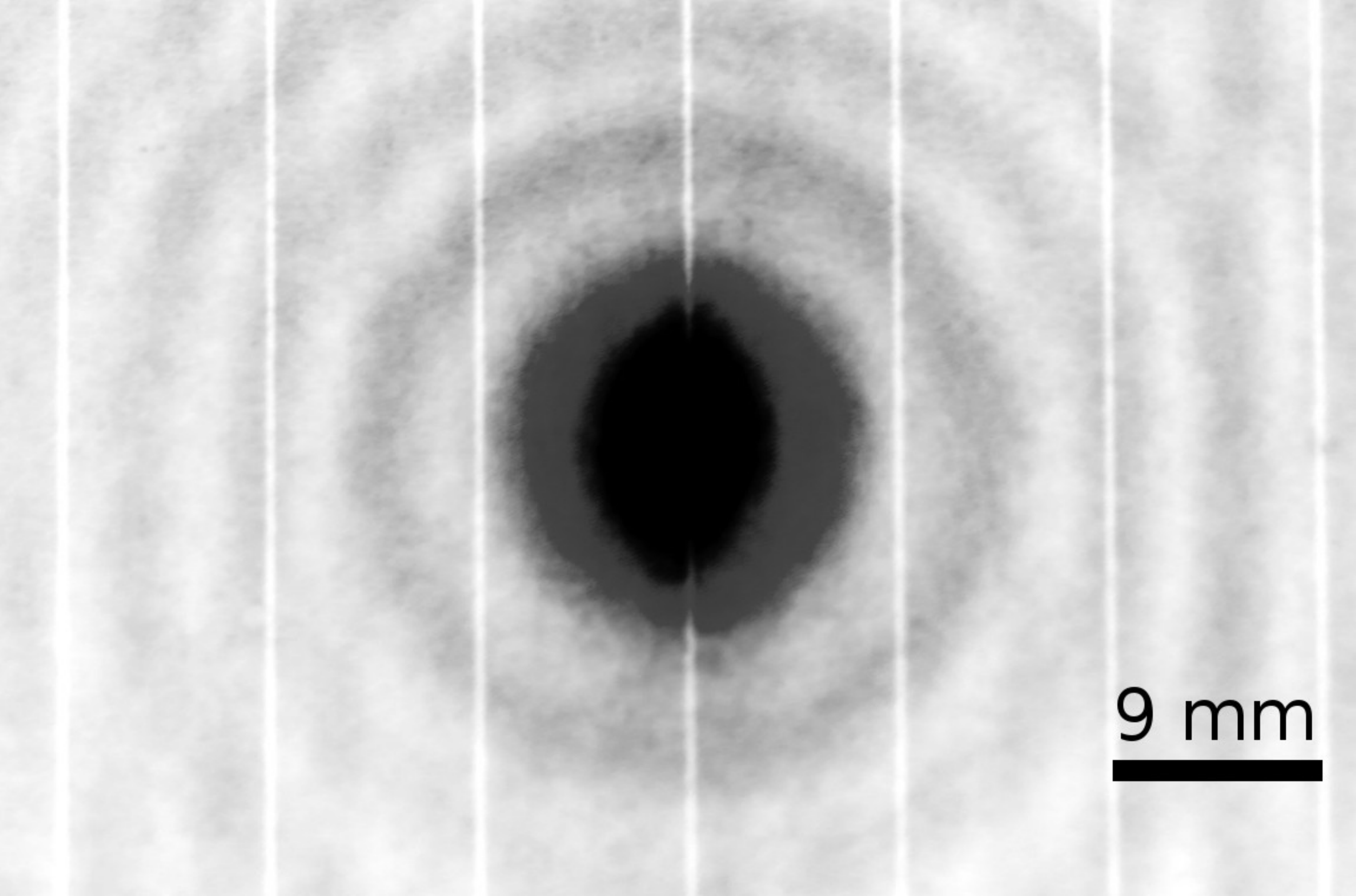}
  \caption{Airy-like pattern produced by the side-pressure laser passing through the trapped particle (inverted colors) and recorded at a distance of 10.5 cm from the trapping spot. The bright central spot is unscattered laser light and the thin vertical stripes are markings on the surface of the screen for scale calibration.}
  \label{fig:airy}
\end{figure}

In order to measure the trapped particle size, we use the ''side-pressure`` beam at low power and record the resulting diffraction pattern (Fig. \ref{fig:airy}). In the geometrical optics regime and for small diffraction angles, the pattern produced by a dielectric sphere is very similar to an Airy pattern, i.e. the diffraction pattern produced by a circular aperture of the same diameter \cite{chen}. The distances between the minima can be then easily linked to a size estimate for the particle \cite{chen}.

\section{Theory}
\subsection{Assumptions and definitions}
The experiment proposed in \cite{sagnac} by Taylor \textit{et al.} assumes the trapped particle to be a Rayleigh scatterer, i.e. a dielectric particle significantly smaller than the wavelength of the trapping light. We generalize their theory to a case of a homogeneous spherical particle of arbitrary size. 

The time dependence of the fields is dropped. A coordinate system is defined for each beam to have $\hat z$ along the beam propagation (with different signs for the CW and CCW beams), $\hat y$ pointing out of the plane in which the interferometer is located (for example, a plane parallel to the optical table) and $\hat x = \hat y \times \hat z$. Additionally, we drop the $y$-component dependence of the fields, i.e. $\phi(x,y) \equiv \phi(x)$ since we investigate only the signal improvement in the $x$-direction. For a compact and clear notation, we describe the propagation of the fields through the system with operators, following the method introduced in \cite{fourier}. For example, to handle the reflections, we define $P$ to be the parity operator such that at every reflection $\phi'(x) = P\phi(x) = \phi(-x)$, where $\phi$ and $\phi'$ are the light fields before and after the reflection respectively. This approach allows us to consider the scattering of incident light by a particle simply as operators acting on the input field. In our case, it is not necessary to know the exact form of these operators, which makes it possible to consider particles of general size.

We assume that the incoming field amplitude distribution $\phi_0$ is symmetric, i.e. $P\phi_0 = \phi_0$, and the field is p-polarized. Additionally, as mentioned before, the backscattered light from the particle is effectively canceled by the polarizer P, so we can ignore it completely in our analysis.

\subsection{Beam propagation}
We define $L_1$ and $L_2$ to be the distances between the beamsplitter and the lens assembly for the CW and CCW beams, respectively. Additionally, $k$ is the wavenumber for the trapping light, defined as $k = 2\pi/\lambda$, with $\lambda$ being the corresponding wavelength. Then, after passing through the beamsplitter and propagating to the lens assembly as shown in Fig. \ref{fig:setup}, the clockwise and the counter-clockwise fields ($\phi_{cw}$ and $\phi_{ccw}$) travel a distance of $L_1$ and $L_2$ respectively. Thus, they become

\begin{align}\label{eq:split}
 \phi'_{cw} &= -r_p\exp({ikL_1}) \phi_0 \notag \\
 \phi'_{ccw} &= t_p f_s \exp({ikL_2}) \phi_0 \: ,
\end{align}
where we used the symmetry of $\phi_0$ and defined field reflection and transmission coefficients $r_p = \sqrt{R_p}$, $t_p = \sqrt{T_p}$. Here $R_p$ and $T_p$ are the intensity reflection and transmission coefficients of the beamsplitter for the p polarization and $f_s=\sqrt{F_s}$ with $F_s$ being the intensity transmission coefficient of the glass plate G for the s polarization. In this and all of the following expressions, the subscripts $s$ and $p$ indicate the polarization for which a given coefficient is applicable. Note that the negative sign before $r_p$ is due to the fact that CW beam is reflected from a hard boundary (the beamsplitter).

After interacting with the particle, the light is scattered both forward and backward, given by the operators $F[x_p]$ and $B[x_p]$, where $x_p$ is a parameter that denotes the position of the particle along the $x$ axis. These operators are defined so that after interacting with the particle, the fields become:

\begin{align}
 &\phi''_{cw} = F[x_p] \phi'_{cw} + B[-x_p] \phi'_{ccw} \notag \\
 &\phi''_{ccw} = F[-x_p] \phi'_{ccw} + B[x_p] \phi'_{cw} \: .
\end{align}
A crucial detail is that the direction of $\hat{x}$ for the CW field is opposite to that of the CCW field, which is the reason for the different signs of $x_p$.

The CW and CCW beams are assumed to pass through $m$ and $n$ mirrors respectively after interacting with the particle. The CW and CCW beams then propagate for an additional distance of $L_2$ and $L_1$ respectively. After the fields join at the beamsplitter, the field in the dark port becomes

\begin{align}\label{eq:raw_raw}
 \phi_d &= \exp(ikL_2) P^{m+1} (f_p r_s F[x_p]\phi'_{cw} + f_s r_p B[-x_p]\phi'_{ccw}) + \notag \\
 &+ \exp(ikL_1) P^{n} (t_s F[-x_p]\phi'_{ccw} + t_p B[x_p]\phi'_{cw}) \: ,
\end{align}
where the exponent $(m+1)$ is due to an additional reflection from the beamsplitter. The coefficients $r_s$, $t_s$ and $f_p$ are defined analogously to Eq. (\ref{eq:split}). Note that, in this case, all the reflections occur \emph{inside} the beamsplitter, so there are no additional sign changes for the fields.

The experiment is set up to ensure equal power in the CW and CCW fields in the trapping point, or equivalently, $r_p = t_p f_s \equiv A$. In addition, the backscattered light is filtered by the polarizer P. Then, after simplification and expansion, Eq. (\ref{eq:raw_raw}) becomes

\begin{equation}\label{eq:raw}
 \phi_d = A \exp(ikL_2) \exp(ikL_1) \left(-r_s f_p P^{m+1} F[x_p] + t_s P^{n}  F[-x_p] \right) \phi_0 \: .
\end{equation}
There, the common exponential phase factor $\exp(ikL_2) \exp(ikL_1)$ is due to the fact that in a Sagnac interferometer, both the CW and CCW beams travel the same distance. Additionally, since the beams are incident at nearly Brewster's angle onto the glass plate G, we can assume that $f_p \approx 1$, so we will drop it in following expressions.

We suppose that the relation $PF[-x_p] = F[x_p]$ is satisfied, which is valid in particular when the particle is a homogeneous sphere and is trapped in the focal plane. Then, Eq. (\ref{eq:raw}) can be simplified to be

\begin{equation}\label{eq:final}
 \phi_d = A \exp\left(ik(L_1+L_2)\right) \left((t_s-r_s)F_s + ((-1)^m r_s - (-1)^n t_s)F_a\right) \phi_0 \: ,
\end{equation}
where, without any loss of generality, we have rewritten the operators in terms of their symmetric and antisymmetric parts: $F[x_p] = F = F_a + F_s$ and $B[x_p] = B = B_a + B_s$. The subscripts indicate symmetry, i.e. $PF_s = F_s$ and $PF_a = -F_a$.

From Eq. (\ref{eq:final}) it can be seen that, in order to obtain constructive interference for the antisymmetric part of the forward scattered field, the total number of mirrors, $(m+n)$, must be odd, which is a general result for the forward scattered field.

In the particular case of a Rayleigh scatterer trapped at an antinode of the standing wave in the trapping plane, as studied by Taylor \textit{et al.} \cite{sagnac}, the forward scattering is the same as the backward scattering. Additionally the polarization is assumed to be the same through the setup, so we can write $r_s=r_p$, $t_s=t_p$, $f_p=f_s=1$. When these hypotheses are applied to the Eq. (\ref{eq:raw_raw}), the resulting expression becomes

\begin{equation}\label{eq:rayleigh-field}
 \phi_{d-ray} = -\exp(2ikL_1) ((t^2 - r^2) F_s + (t+r)^2 F_a) \phi_0 \: ,
\end{equation}
which corresponds to the result of Taylor \textit{et al.} \cite{sagnac}, thus confirming our approach.

\subsection{Signal enhancement}\label{sec:improvement}
The signal-to-noise ratio is typically defined as the ratio between the variance of the signal and the variance of the noise:

\begin{equation}
 SNR = \frac{\Var(i_{signal})}{\Var(i_{noise})} \: .
\end{equation}
Following Taylor \textit{et al.} \cite{sagnac}, we notice that the laser powers often used in optical trapping experiments exceed greatly the saturation threshold of commercial QPD detectors. In our case, we use the a QP50-6 QPD by First Sensor which has a saturation threshold of approximately 1 mW. Therefore, the light field in the dark port has to be attenuated to an average power level slightly below the saturation threshold. This level has to be such as to avoid saturating the detector with the signal maxima. Then, the comparison between Sagnac BFP and traditional BFP should be performed by assuming equal average power incident on the QPD, which is equivalent to:

\begin{equation}
 \overline{i_{background-bfp}} = \overline{i_{background-sagnac}} \equiv \overline{i_{background}} \: ,
\end{equation}
where we used the assumption that the average position signal level is always zero, i.e.

\begin{equation}
 \overline{i_{total}} = \overline{i_{background}} + \overline{i_{signal}} = \overline{i_{background}} \: .
\end{equation}

Since the optical shot noise is a function of average light power incident on the detector, and the average light power in our comparison is the same both for Sagnac and traditional BFP, we have

\begin{align}
 SNR_{sagnac} = \frac{\Var(i_{signal-sagnac})}{\Var(i_{noise})} \: , \notag \\
 SNR_{bfp} = \frac{\Var(i_{signal-bfp})}{\Var(i_{noise})} \: ,
\end{align}
where $i_{noise}$ is the intensity signal associated with the noise, while $i_{signal-sagnac}$ and $i_{signal-bfp}$ are, respectively, intensity signals associated with the particle position.

In our case, it will be useful to define the signal-to-background ratio (SBR) as follows:
\begin{equation}
  SBR^2 = \frac{\Var(i_{signal})}{\overline{i_{background}}^2} \: ,
\end{equation}
which is equivalent to:

\begin{equation}
 \Var(i_{signal}) = (SBR \cdot \overline{i_{background}})^2 \: .
\end{equation}
Finally, we obtain

\begin{equation}\label{eq:snr-sbr}
 \frac{SNR_{sagnac}}{SNR_{bfp}} = \left( \frac{SBR_{sagnac}}{SBR_{bfp}} \right)^2 \: .
\end{equation}
Therefore, the measurements that presented in this work can be expressed equivalently as the signal-to-noise ratio (SNR) improvement or the signal-to-background ratio (SBR) improvement.

The signal amplitude and the background level are calculated by setting $F_s \phi_0 = \phi_s$ and $F_a \phi_0 = \phi_a$ and by assuming $(m+n)$ to be odd. Then Eq. (\ref{eq:final}) becomes

\begin{equation}
 \phi_d(x) = A((t_s - r_s)\phi_s(x) + (t_s + r_s)\phi_a(x)) \: ,
\end{equation}
up to a complex multiplier of unit modulus.

In order to calculate the signal-to-background ratio improvement, it is not necessary to integrate the intensity over $x$. Instead, we note the fact that the sum of intensities on a symmetrical pair of points $\{-x, x\}$, i.e. $|\phi_d(x)|^2 + |\phi_d(-x)|^2$, is proportional to the total light power incident onto the QPD, and is given by

\begin{equation}\label{eq:sum}
 |\phi_d(x)|^2 + |\phi_d(-x)|^2 = 2((t_s - r_s)^2 |\phi_s(x)|^2 + (t_s+r_s)^2 |\phi_a(x)|^2) \approx 2(t_s - r_s)^2 |\phi_s(x)|^2 \: ,
\end{equation}
where we used the experimentally valid hypothesis of sufficiently small oscillations of the particle, i.e. $|\phi_a(x)|^2 \ll |\phi_s(x)|^2$.

On the other hand, the ``X-signal'' in the QPD detector is in general proportional to the difference of the intensities

\begin{equation}
 |\phi_d(x)|^2 - |\phi_d(-x)|^2 = 4(T_s-R_s) Re(\phi_s(x) \phi_a(x)^*) \: .
\end{equation}
In our case of $|\phi_a(x)|^2 \ll |\phi_s(x)|^2$, we can use the right-hand side of the Eq. (\ref{eq:sum}). Then, the signal-to-background (SBR) ratio ($k_{sbr-sagnac}$) becomes

\begin{equation}\label{eq:stb}
 k_{sbr-sagnac} = \frac{|\phi_d(x)|^2 - |\phi_d(-x)|^2}{|\phi_d(x)|^2 + |\phi_d(-x)|^2} = 2 \cdot \frac{t_s+r_s}{t_s-r_s} \cdot \frac{Re(\phi_s(x) \phi_a(x)^*)}{|\phi_s(x)|^2} \: .
\end{equation}
The degenerate case when $t_s=t_p=1$ and $r_s=r_p=0$ (i.e. no interferometer) corresponds to typical back-focal-plane interferometry. The signal-to-background ratio ($k_{sbr-bfp}$) is then given by

\begin{equation}
 k_{sbr-bfp} = \frac{|\phi_d(x)|^2 - |\phi_d(-x)|^2}{|\phi_d(x)|^2 + |\phi_d(-x)|^2} = 2 \cdot \frac{Re(\phi_s(x) \phi_a(x)^*)}{|\phi_s(x)|^2} \: .
\end{equation}
Therefore, the use of a Sagnac interferometer allows to improve the SBR over the typical back-focal-plane interferometry by a factor of

\begin{equation}\label{eq:improvement}
 \frac{k_{sbr-sagnac}}{k_{sbr-bfp}} = \frac{t_s+r_s}{t_s-r_s} \: ,
\end{equation}
which can be linked with Eq. (\ref{eq:snr-sbr}) to a SNR improvement of

\begin{equation}
 \frac{SNR_{sagnac}}{SNR_{bfp}} = \left(\frac{k_{sbr-sagnac}}{k_{sbr-bfp}} \right)^2 = \left( \frac{t_s+r_s}{t_s-r_s} \right)^2 \: .
\end{equation}
In our experimental case, the expression (\ref{eq:improvement}) gives a maximum achievable signal-to-background ratio improvement factor of $\approx 11.9$ given the transmission and reflection coefficients of the Thorlabs BSW26 beamsplitter ($R_s \approx 56.47\%$, $T_s = 40.68\%$). This corresponds to a maximum signal-to-noise ratio improvement factor of $\approx 140$.

As a verification of consistency, we may apply the above procedure to Eq. (\ref{eq:rayleigh-field}), which is equivalent to calculating the SNR improvement for the experimental case of Taylor \textit{et al.} \cite{sagnac}, and thus obtain

\begin{align}
 |\phi_{d-ray}(x)|^2 + |\phi_{d-ray}(-x)|^2 &= 2((t^2 - r^2)^2 |\phi_s(x)|^2 + (t+r)^4 |\phi_a(x)|^2) \approx 2(t^2 - r^2)^2 |\phi_s(x)|^2 \\ \notag
 |\phi_{d-ray}(x)|^2 - |\phi_{d-ray}(-x)|^2 &= 4(t^2 - r^2) (t + r)^2 Re(\phi_s(x) \phi_a(x)^*) \: .
\end{align}
Then, analogously with Eq. (\ref{eq:stb}), we obtain the signal-to-background ratio for the experimental case of Rayleigh scatterers in \cite{sagnac}:

\begin{equation}
 k_{sbr-sagnac-ray} = 2 \cdot \frac{(t+r)^2}{(t-r)^2} \cdot \frac{Re(\phi_s(x) \phi_a(x)^*)}{|\phi_s|^2} \: .
\end{equation}
Finally, the signal-to-noise ratio improvement can be computed in analogy with Eqs. (\ref{eq:improvement}) and (\ref{eq:snr-sbr}):

\begin{equation}
 \frac{SNR_{sagnac-ray}}{SNR_{bfp-ray}} = \left(\frac{k_{sbr-sagnac-ray}}{k_{sbr-bfp-ray}} \right)^2 = \left( \frac{(t+r)^2}{t^2-r^2} \right)^2 = \frac{(\sqrt{T}+\sqrt{R})^4}{(T - R)^2}\: ,
\end{equation}
which is exactly the result obtained by Taylor \textit{et al.} \cite{sagnac}.

\section{Results and discussion}
\subsection{Particle generation and levitator-assisted trapping}
The inkjet cartridge-based particle generation with evaporative diameter reduction proved to be reliable over several months of operation. Due to the abundance of nozzles (over 50) in the cartridge \cite{inkjet}, problems with a non-working nozzle, e.g. due to clogging, were easily solved by using an alternative nozzle. By using a mixture of 5\% glycerol in water, stable droplet generation of particles with a constant diameter of $11 \pm 1$ $\mu$m (average over 5 measurements) was achieved. Their sizes were measured by using diffraction patterns similar to the one shown in Fig. \ref{fig:airy}, as described in section \ref{sec:diameter}.

The levitation-assisted delivery greatly simplified the trapping of droplets into the horizontal counterpropagating optical tweezers. The illumination of the whole stream of falling droplets, as in Fig. \ref{fig:cell_real}, provided us with a visual feedback on the cartridge's orientation during fine adjustments. Normally, 30-50 droplets had to be generated to finely adjust the cartridge orientation in order to compensate for mechanical drifts. After this adjustment all the subsequent droplets descended into the levitator trap where they were trapped. By eliminating the remaining long-term mechanical drifts, it should be possible to achieve automatic ``trap on demand'' operation with single droplets.

\subsection{Position measurement and improvement of signal-to-background ratio}

\begin{figure}[htpb]
\centering  
  \includegraphics[width=0.7\textwidth]{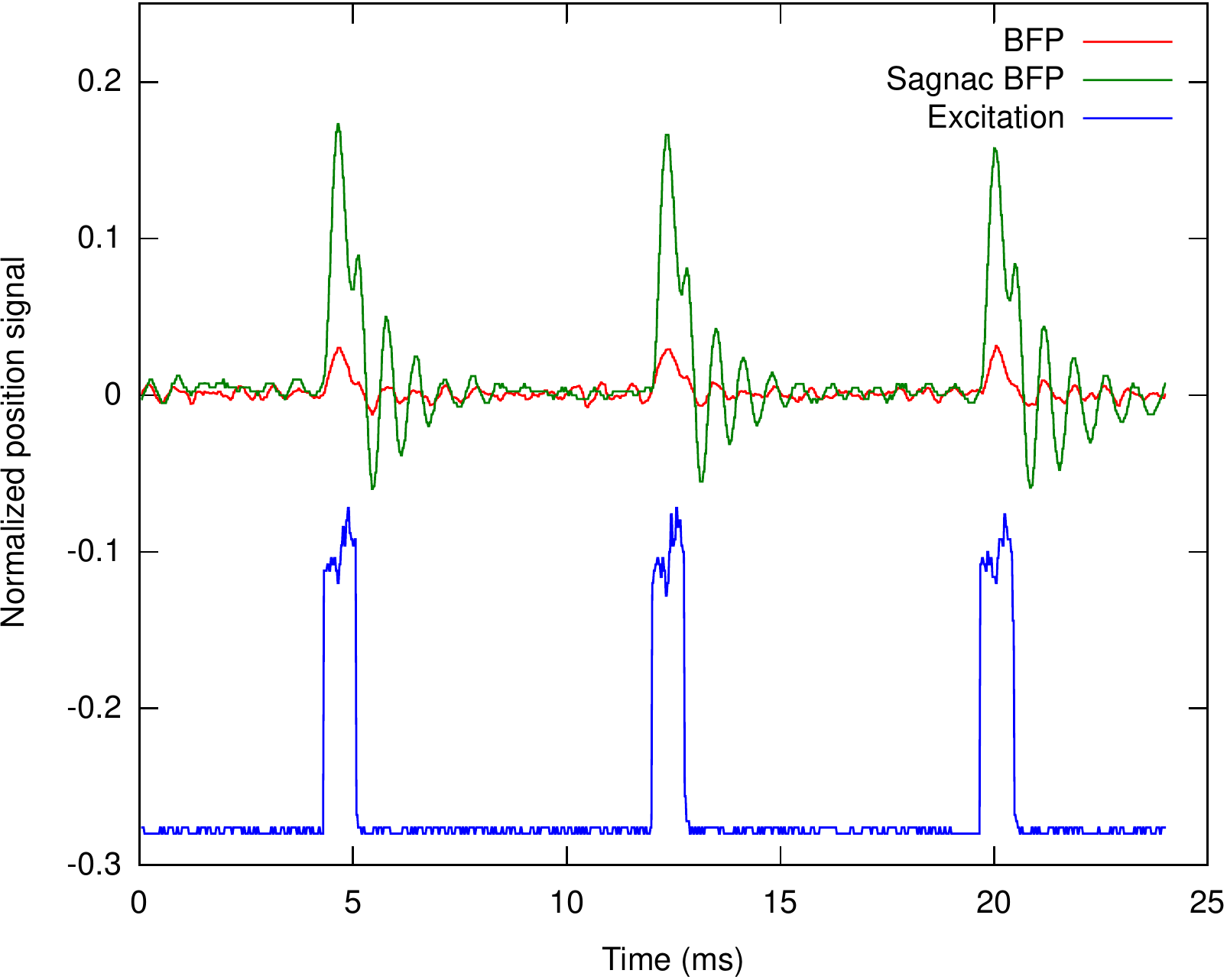}
  \caption{Comparison between the normalized position signals obtained with the traditional BFP interferometry (red) and Sagnac-enhanced BFP interferometry (green). The side-pressure laser power (blue, arbitrary units) is driven by a 10\% duty cycle, 130 Hz TTL wave.}
  \label{fig:data}
\end{figure}

The signal that we obtained was normalized to be consistent with Eq. (\ref{eq:stb}), i.e. is the ratio of the signals obtained from the QPD detector
\begin{equation}\label{eq:norm}
 x_{norm} = \frac{X}{SUM} \: ,
\end{equation}
where $X$ is the difference of power between the left and right halves of the QPD detector and $SUM$ is the total power arriving at the detector.
The data expressed in this way is effectively the signal divided by the background and thus, can be used directly to compare the typical BFP measurements to the Sagnac-enhanced BFP measurements.

Figure \ref{fig:data} shows typical curves of the normalized position signals, as specified by the Eq. (\ref{eq:norm}). The side pressure laser is applied with a frequency of 130 Hz and duty cycle of 10\%. The red curve in the figure displays the measured normalized position using the BFP interferometry whereas the green curve displays the same quantity using Sagnac-enhanced BFP interferometry. 

The particle position signal shows a typical response of an underdamped harmonic oscillator to a periodic rectangular-wave external force (blue curve in Fig. \ref{fig:data}). This is expected as the damping ratio is much less that one for a 11-micrometer particle in air trapped in optical tweezers with a total laser power of 1W. The small structure during the laser-on period, shown as two small bumps in the power signal (blue) is due to mode-jumping in the side-pressure laser during the modulation. These power fluctuations give rise to similar bumps in the particle position (green and red).

Note the fact that the oscillations in the normalized ``typical BFP'' signal, as shown in Fig. \ref{fig:data}, are much smaller than  unity. This justifies our use of the small-oscillations approximation in the section \ref{sec:improvement}.

After acquiring the Sagnac-enhanced signal together with its corresponding ``typical BFP'' signal, we divided the Sagnac-enhanced waveforms over the ``typical'' BFP waveform. The resulting dataset represents the ratio between the two signals at every timestep. Assuming this ratio to be constant and considering the uncertainty associated with our acquisition system, we calculated a SNR improvement of $5.6 \pm 0.1$, which is consistent with the theory shown in section \ref{sec:improvement}. The theoretical limit for the improvement, as calculated in section \ref{sec:improvement}, is approximately 11.9. In practice, this value will be reduced due to the imperfections of the setup, such as the sub-optimal performance of our aspherical lenses at a wavelength of 532 nm and the imperfect mode matching in the interferometer. The mode-matching can be suboptimal if, for example, the overlap between the CW and CCW beams is not complete or in the case where the beams are not ideally centered on the aspherical lenses. 

Finally, compared to typical back focal plane interferometry, the signal-to-shot-noise improvement corresponding to our measurements is more than 30-fold according to Eq. (\ref{eq:snr-sbr}).

\section{Conclusion}
We have extended the theory presented by Taylor \textit{et al.} \cite{sagnac} to a case of arbitrarily-sized spherical particles trapped in orthogonally-polarized counter-propagating optical tweezers. We experimentally demonstrated a significant improvement in the signal-to-background ratio for position detection of 11 $\mu$m-sized particles by embedding counterpropagating optical tweezers into a Sagnac interferometer. The results are consistent with our theoretical prediction, thus validating our approach. The obtained data indicate an increase of signal-to-shot-noise ratio of more than 30 times compared to typical back focal plane interferometry.

We have constructed a system based on a low-cost commercial inkjet cartridge \cite{inkjet} and an optical levitator \cite{levitator-feeding} that could efficiently generate, trap and guide into the optical tweezers single glycerol droplets in air, while having precise evaporation-based control of droplet sizes.  By eliminating the remaining long-term mechanical drifts in this system, it should be possible to achieve automatic ``trap on demand'' operation with single droplets.

We hope that our results will be useful for experiments requiring improved resolution in BFP interferometry position detection for spherical particles in the Mie and geometrical optics regimes, such as the counterpropagating setup for center-of-mass motion cooling of 3 $\mu$m particles shown by Li \textit{et al.} in \cite{cooling}.

\section*{Acknowledgments}
We thank José Luis Meza for his invaluable assistance during the development of the project in the Faculty of Sciences, UNAM. This work was supported by the Carl Trygger Foundation for Scientific Research (contract CTS13:169), The Swedish Council for Higher Education through the Linnaeus-Palme International Exchange Program (contract 4332-2013) and by the grant ”Bottlenecks for particle growth in turbulent aerosols” from the Knut and Alice Wallenberg Foundation (Dnr. KAW 2014.0048).

\bibliographystyle{plain}
\bibliography{bib}{}

\end{document}